\documentclass[1ps,preprint]{mn2e}
\usepackage{graphicx}
\usepackage{longtable}
\usepackage{pdflscape}
\usepackage{rotating}
\newcommand{\sop}{Solar Physics}
\newcommand{\nimpb}{Nucl. Inst. and Meth. in Phys. Research B}
\newcommand{\araa}{Annu. Rev. Astron. Astrophys.}
\newcommand{\adndt}{Atom. Data and Nucl. Data Tables}
\title[Soft X-ray emissions of highly charged Si ions ...]{Soft X-ray emissions of highly charged Si~VII--Si~XII in cool star--Procyon}
\author[G. Y. Liang et al.]{G. Y. Liang$^{\star}$ and G. Zhao\thanks{Corresponding
author:~~G.Y. Liang~~(gyliang@bao.ac.cn);  ~~G.
Zhao~~(gzhao@bao.ac.cn)}\\
       National Astronomical Observatories, Chinese Academy of Sciences, Beijing
 100012, P. R. China}
\begin{document}
\date{Received date  / Accepted date }
\pagerange{\pageref{firstpage}--\pageref{lastpage}} \pubyear{2007}

\maketitle

\label{firstpage}
\begin{abstract}
Different observation data for cool star---Procyon (Obs\_IDs of
63, 1461 and 1224) available from {\it Chandra Data Public
Archive} were co-added and analyzed. Emissivities of emission
lines of highly charged silicon ions (Si~VII--Si~XII) were
calculated over temperatures by adopting the published data of
Liang et al. (2007, {\it Atom. Data and Nucl. Data Tables}, {\bf
93}, 375). Using the emission measure derived by Raassen et al.
(2002, A\&A, {\bf 389}, 228), the theoretical line fluxes are
predicted, and the theoretical spectra are constructed by assuming
the Gaussian profile with instrumental broadening (0.06~\AA\,). By
detailed comparison between observation and predictions, several
emissions lines are identified firstly such as emissions at
43.663~\AA\, (Si~XI), 45.550~\AA\, (Si~XII), 46.179~\AA\,
(Si~VIII), 50.874~\AA\, (Si~X), 64.668~\AA\, (Si~IX), and
73.189~\AA\, (Si~VII) etc. Several emission lines are re-assigned
in this work, such as the emission line at 52.594~\AA\, to Si~X
(52.612~\AA\,), at 69.641~\AA\, to the blending of Si~VII
(69.632~\AA\,) and Si~VIII (69.664~\AA\,) lines, as well as at
70.050~\AA\, to Si~VII (70.027~\AA\,). The prediction reveals the
large discrepancies between the $3s$--$2p$ line (63.715~\AA\,)
{\it versus} $3d$--$2p$ line (61.012~\AA\,) again for lower charge
stage Si~VIII. Solar flare observation is also added for the
assessment of present calculation. Different assignments for some
lines between Procyon and solar flare, have been found to be due
to the hotter emitter in Sun than in Procyon coronae.
\end{abstract}

\newpage
\section{Introduction}
Since the launch of {\it Chandra} and {\it XMM}-Newton X-ray
missions, a large quantity of high quality X-ray spectra with
high-resolution has been obtained for nearly all classes of
astrophysical X-ray
sources~\cite{BGK00,CHD00,ABG01,BBG01,NMS01,RBK01,FCD04}, which
allows detailed plasma diagnostics to be performed for a wide
range of celestial objects.

In the spectra of cool stars, rich emission lines of highly
charged iron ions were detected due to its high abundance and high
effective collecting area (200~cm$^2$ at 8.5~\AA\, for
HEG/MEG+ACIS-S instrument; 35~cm$^2$ at 10.0~\AA\, for LETG+HRC-S
instrument\footnote{http://cxc.harvard.edu/ciao/manuals.html})
between 6---18~\AA\,. In this wavelength region, a good agreement
between predictions and observations is achieved so
far~\cite{BCK01}, which profits from the Iron-Project~\cite{HBE93}
and earlier running program--Opacity project~\cite{SYM94}.
Moreover, many laboratory experiments have been performed for iron
by adopting grating and crystal spectrometers with
high-resolution. The laboratory platform can be early vacuum
spark, tokamak used for magnetic confinement fusion study, intense
laser used for inertial confinement fusion research, and newly
developed facility--electron beam ion trap (EBIT, Beiersdorfer
2003). Though the difference of $3s$--$2p$ line intensity {\it
versus} $3d$--$2p$ line intensity of Fe~XVII is a debating
problem, a good theoretical performance is obtained when compared
with spectra of highly charged sulfur and argon, as revealed by
experimental measurements at Livermore EBIT~\cite{LBB03,LBB05a}.

Besides rich emission lines of highly charged Fe ions, some
emission lines of L-shell calcium, argon, sulfur and silicon have
been identified for cool stars. Yet detailed analysis for spectra
of these ions receive a scant attention due to the low effective
area over the wavelength range of 35---100~\AA\, spanned by lines
of L-shell Si, Ar and Ca ions, and the absence of accurate atomic
data of these ions. Several
literatures~\cite{KPW98,DKR99,KST00,LZ06,LZS06c} also reveal the
great potentials of L-shell silicon ions. Such as, the line
intensity ratio $I$(50.524)/$I$(50.691) of Si~X,
$I$(52.306)/$I$(46.391) of Si~XI are good diagnostic methods for
the electron density; the line ratio $I$(52.306)/$I$(43.743) of
Si~XI shows a good performance for the electron temperature
($T_{\rm e}$) determination. Similar characteristics were explored
for S~X, Ar~XIV and Ca XVI
etc.~\shortcite{KCF93,KPM00,KAW01,KAK03}. Furthermore, spatial
information of coronae could be assessed indirectly using
correlation ($EM=n_e^2V$) of the electron density with emission
measure.

A laboratory measurement for silicon has been performed at the
Livermore EBIT-II. Unfortunately, only the spectra between
80---90~\AA\, is shown~\cite{LBB05b}, yet it is enough to
demonstrate the large differences between the measurement and
theory. Recently, accurate atomic data including energy levels and
spontaneous radiative decay rates were calculated for
Si~IX---Si~XI by consideration of large configuration interaction
and relativistic effects~\cite{LZZ07}. A self-consistent electron
impact excitation rates over a large temperature grids are
available, which are based on relativistic distort-wave (RDW)
method. A laboratory measurement of soft X-ray spectra
(40---180~\AA\,) of highly charge silicon ions were also performed
by irradiating the silicon target by intense laser beam at
Institute of Physics, Chinese Academy of Sciences, which
satisfactorily benchmarks the theoretical modelling at high
density~\cite{LZZL07}.

In this work, we calculate the line emissivities of highly charged
silicon ions (Si~VI---Si~XII) by adopting new available and
accurate atomic data. The available observation data for cool
star--Procyon (Obs\_IDs 63, 1461 and 1224) are co-added and
analyzed. By detailed comparison between the predictions and the
co-added spectrum. Emissions lines of highly charged silicon are
investigated. The observation of solar flare is also summarized
for the assessment of the calculation.

\section{Observations and Data analyses}
Procyon (F5~IV--V) is a solar-like star at a distance of 3.5~pc
with mass of 1.75$M_{\rm \sun}$ and radius of 2.1$R_{\rm \sun}$,
which has been observed by every X-ray space missions such as {\it
Chandra} and {\it XMM}-Newton, for calibration. From {\it Chandra}
Public Data Archive\footnote{http://cxc.harvard.edu/cda/}, four
observations for Procyon can be available, with one observed by
HRC-I instrument and none grating. For other three observations,
HRC-S instrument was used in combination with low energy
transmission grating (LETG), which covers a wavelength range of
6--176~\AA\,. The description of the data sets is listed in Table
1.
\begin{table}
\centering
    \caption[I]{Observation data with HRC-S/LETG instrument for Procyon from {\it Chandra} public data archive.}
    \vspace{0.2cm}
      \[
      \begin{array}[h]{ccccc} \hline\hline
{\rm Seq.} & {\rm Obs\_ID} & {\rm Instr./} & {\rm Exposure} & {\rm Start} \\
{\rm Num.}     &               & {\rm Grating} & {\rm time~(ks)}    & {\rm time} \\
 \hline
280174 & 1224 & {\rm HRC-S/LETG} & 20.93 & 1999-11-08 \\
280411 & 1461 & {\rm HRC-S/LETG} & 70.25 & 1999-11-07 \\
290032 &  63  & {\rm HRC-S/LETG} & 70.15 & 1999-11-06 \\
\hline
         \end{array}
      \]
      \end{table}

In this work, the reduction of the data sets uses {\it Sherpa}
software package in {\sf CIAO}, version 3.3, with the science
threads for HCR-S/LETG observations. The three spectra (with
Obs\_IDs of 63, 1461 and 1224) are co-added with the {\sf
add\_grating\_spectra} tool to improve the signal-to-noise ratio,
which results in a spectrum with total exposure time of 159.5~ks
after times of bad counts were excluded. Similarly, the associate
auxiliary response files (ARFs) are averaged by this tool. Here,
the positive and negative spectra are analyzed separately by
consideration of the different chip gaps for the two diffraction
orders. Figs.~1--3 [histogram curves] show the co-added spectra
for positive and negative diffraction orders in wavelength range
of 43---88~\AA\, spanned by emission lines of highly charged
silicon ions (Si~VI---Si~XII).

Line fluxes are determined by modelling the spectra locally with
narrow Gaussian profiles and constant value representing
background and (pseudo-)continuum emissions determined in
line-free region. The observed line width is about 0.06~\AA\, over
the interested region, which is comparable with the broadening of
instrument for point-like source. The fluxes have been obtained
after correction for the effective area. In the fitting, 1$\sigma$
uncertainty was adopted to determine the statistical errors for
the line fluxes. Here, only the line fluxes of highly charged
silicon ions are listed in Table 2. For comparison, the line
fluxes from Raassen et al.~\shortcite{RMA02} are listed, which
reveals a good agreement for most peaks, although different
observations are used.

Additionally, solar observation data in this wavelength region are
also listed for completeness. The wavelength and line intensity
(in unit of photons$\cdot$cm$^{-2}$s$^{-1}$~acrsec$^{-1}$) are
from work of Acton et al.~\shortcite{ABB85}, whereas the line
intensities being less than
10~photons~cm$^{-2}$s$^{-1}$~acrsec$^{-1}$ have not been given.

\section{Theoretical models}
Collisional-radiative (CR) models for highly charged
Si~VII---Si~XII ions are constructed based on the accurate atomic
data from work of Liang et al.~\shortcite{LZZ07} and their
unpublished data. These data are generated with the Flexible
Atomic Code (FAC) provided by Gu~\shortcite{Gu03}. 878, 312, 560,
320 and 350 energy levels have been included in predictions of
line emissivities of Si~VII---Si~XI, respectively. These levels
belongs to not only singly excited configurations, but also some
doubly excited configurations for accounting for configuration
interaction as fully as possible. Some energy levels are replaced
by available experimental values from the National Institute of
Science and Technology (NIST)
database\footnote{http://physics.nist.gov/PhysRefData/ASD/levels\_form.html}.
For Si~XII, 40 energy levels are available from Chianti
database~\cite{LZY06}, and used here. All possible decay channels
among above listed levels by E1, M1, E2 and M2 type transitions,
have been included in the present model for each charge states.

A self-consistent calculation of electron impact excitation has
been performed for the five ions (Si~VII--Si~XI), which based on
the RDW method~\cite{Gu03}. These data has been assessed for
Si~IX--Si~XI in our previous work~\cite{LZZ07}. For $\Delta n=0$
transitions among levels of ground and lower excited
configurations, the excitation data is replaced by available
$R-$matrix data, which properly considers resonant effects in the
threshold region. To our best knowledge, the $R-$matrix
calculation is not available for Si~VII. So present RDW
calculation of electron impaction excitation is used for Si~VII by
further consideration of self-consistency. Bell et
al.~\shortcite{BMR01} calculate the electron impact excitation by
using {\it ab initio R-}matrix method for excitations among levels
of ground configuration $1s^22s^22p^3$ of Si~VIII. These data
replaces our results in predictions of line emissivities. For
Si~IX, resonant effects among levels of ground configuration, have
been considered in predictions of level populations by using the
data of Aggarwal~\shortcite{Agg83}. Zhang et al.~\shortcite{ZGP94}
performed $R-$matrix calculation for 105 transitions among the 15
fine-structure energy levels belongings to $2s^22p$, $2s2p^2$ and
$2p^3$ configurations of Si~X. However, an error was noted by
Keenan~\shortcite{Kee00}, who re-calculated the electron impact
excitation rates for the 105 transitions with the $R-$matrix
method. These data is adopted here for Si~X. $R-$matrix results
for 29 independent transitions among 10 fine-structure energy
levels belongings to $2s^2$, $2s2p$ and $2p^2$ configurations of
Si~XI were presented by Berrington et al.~\shortcite{BBD85}. These
accurate data replace the RDW calculations in the predictions of
level populations of each charge states.

\section{Results and discussions}
By adopting the atomic data described in above section, line
emissivities of Si~VII--Si~XII are calculated at an electron
density of 3.0$\times10^8$~cm$^{-3}$ and electron temperatures
ranging 0.1--5~MK. The electron density is a typical value for
cool star with lower activities as revealed by previous
works~\cite{NSB02,LZ06}. The silicon abundance adopts solar
photospheric value, while the ionization equilibrium uses the
result of Mazzotta et al.~\shortcite{MMC98}. Previous works have
revealed that the emission measure ($EM=n_{\rm e}n_{\rm H}V$) of
Procyon has a continuous distribution and mainly dominates around
1--3~MK [see Fig.~3 in {\it Ref.} Raassen et
al.~\shortcite{RMA02}]. The theoretical line fluxes of highly
charged silicon ions, are derived by combining the line
emissivities and the emission measure. The predicted line fluxes
are listed along with the observed values for comparison in Table
2. Furthermore, the theoretical line fluxes couple the effective
area extracted from ARFs files, and are folded by Gaussian
profiles with full-width at half-maximum (FWHM) of 0.06~\AA\,
representing the observed line width. Figs.~1--3 show the
theoretical spectra by color curves for different charge states.
In these figures, the positive and negative diffraction are
analyzed, separately. For most strong emission lines, the
theoretical calculations agree well with the observations within
the statistical uncertainty.

In solar case, we calculate the line intensity at a higher
electron density of 5.0$\times10^9$~cm$^{-3}$ and temperatures of
peak fractions for each charge state in the ionization
equilibrium~\cite{MMC98}. Further the predictions are scaled by
solar values at 44.20~\AA\, (Si~XII), 46.40~\AA\, (Si~XI),
50.52~\AA\, (Si~X), 55.41~\AA\, (Si~IX), 61.09~\AA\, (Si~VIII) and
70.05~\AA\, (Si~VII) for each charge state, respectively. In the
following, we discuss the calculation and comparison in sequence
of charge states.
\begin{table*}
\centering
    \caption[I]{The wavelengths and line fluxes (in unit of 1.0$\times10^{-4}$~phot.~cm$^{-2}$s$^{-1}$) of
    emission lines of Si~VI--Si~XII in Procyon. The column of RMS02 denotes the value from work of Raassen et
    al.~\shortcite{RMA02}. Wavelengths ($\lambda_{\sun}$) and fluxes ($F_{\sun}$) in solar flare observation are
    from work of Acton et al.~\shortcite{ABB85}.
    The flux with subscript of 3CIE denotes the predicted fluxes from the present emissivity
    and the 3-temperature model~\cite{RMA02} for the Procyon coronae. $F_{\sun}^p$ denotes the predicted
    line intensities normalized by solar values at 44.20~\AA\, (Si~XII), 46.40~\AA\,
(Si~XI), 50.52~\AA\, (Si~X), 55.41~\AA\, (Si~IX), 61.09~\AA\,
(Si~VIII) and 70.05~\AA\, (Si~VII) lines for each charge state,
respectively.}
    \vspace{-0.2cm}
      \[\hspace{-0.7cm}
      \begin{array}[h]{lccccccclcccllll} \hline\hline
         & & & & & & & & & \multicolumn{3}{c}{\rm This~work} & \multicolumn{4}{c}{\rm Transitions} \\
{\rm ID} & {\rm \lambda_{-}~(\AA)} & {\rm Flux_{-}} & {\rm
\lambda_{+}~(\AA)} & {\rm Flux_{+}} & {\rm RMA02} & {\rm
\lambda_{\sun}} &  F_{\rm \sun} & {\rm Ions} & {\rm \lambda~(\AA)}
& F_{\tiny 3CIE} & F^p_{\tiny \sun}& \multicolumn{2}{c}{\rm
Upper~level} & \multicolumn{2}{c}{\rm Lower~level} \\ \hline
1  & 43.663 & 0.10(9)  & 43.656 & 0.11(7)  &          & 43.65 & 21 & {\rm Si~XI  } & 43.683 & 0.08 & 7  & 2s3p            & ^3P_1     & 2s^2      & ^1S_0     \\
2  & 43.753 & 0.61(9)  & 43.731 & 0.55(7)  & 0.54(8)  & 43.75 & 90 & {\rm Si~XI  } & 43.763 & 0.39 & 35 & 2s3p            & ^1P_1     & 2s^2      & ^1S_0     \\
3  & 44.029 & 0.49(8)  & 44.002 & 0.33(6)  & 0.43(8)  & 44.02 & 138& {\rm Si~XII } & 44.019 & 0.43 & 125& 3d              & ^2D_{3/2} & 2p        & ^2P_{1/2} \\
4  & 44.183 & 0.73(9)  & 44.131 & 0.60(7)  & 0.67(10) & 44.16 & 224& {\rm Si~XII } & 44.165 & 0.76 & 224& 3d              & ^2D_{5/2} & 2p        & ^2P_{3/2} \\
5a & 44.275 & 0.28(8)  & 44.193 & 0.64(7)  & 0.52(10) & 44.20 & 292& {\rm Si~IX  } & 44.213 & 0.08 & 3  & 2s^22p4d        & ^3D_1     & 2s^22p^2  & ^3P_2     \\
5b &        &          &        &          &          &       &    & {\rm Si~IX  } & 44.215 & 0.10 & 4  & 2s^22p4d        & ^3D_2     & 2s^22p^2  & ^3P_1     \\
5c &        &          &        &          &          &       &    & {\rm Si~IX  } & 44.249 & 0.15 & 6  & 2s^22p4d        & ^3D_3     & 2s^22p^2  & ^3P_2     \\
6  & {\bf 45.550} & 0.14(8)  & 45.490 & 0.12(6)  &          & 45.51 & 57 & {\rm Si~XII } & 45.521 & 0.22 & 67 & 3s              & ^2S_{1/2} & 2p        & ^2P_{1/2} \\
7  & 45.694 & 0.34(9)  & 45.664 & 0.22(7)  & 0.20(4)  & 45.68 & 109& {\rm Si~XII } & 45.691 & 0.45 & 135& 3s              & ^2S_{1/2} & 2p        & ^2P_{3/2} \\
8a & {\bf 46.179} & 0.16(5)  & 46.110 & 0.25(6)  &          &       &    & {\rm Si~VIII} & 46.133 & 0.09 & 1  & 2s^22p^2(^3P)5d & ^4P_{3/2} & 2s^22p^3  & ^4S_{3/2} \\
8b &        &          &        &          &          &       &    & {\rm Si~VIII} & 46.148 & 0.05 & 1  & 2s^22p^2(^3P)5d & ^4P_{5/2} & 2s^22p^3  & ^4S_{3/2} \\
9a & 46.301 & 0.31(5)  & 46.272 & 0.31(6)  & 0.25(7)  & 46.30 & 55 & {\rm Si~XI  } & 46.263 & 0.08 & 8  & 2s3d            & ^3D_1     & 2s2p      & ^3P_0     \\
9b &        &          &        &          &          &       &    & {\rm Si~XI  } & 46.298 & 0.15 & 15 & 2s3d            & ^3D_2     & 2s2p      & ^3P_1     \\
9c &        &          &        &          &          &       &    & {\rm Si~XI  } & 46.313 & 0.06 & 6  & 2s3d            & ^3D_1     & 2s2p      & ^3P_1     \\
10 & 46.407 & 0.49(5)  & 46.366 & 0.42(6)  & 0.40(8)  & 46.40 & 50 & {\rm Si~XI  } & 46.399 & 0.41 & 50 & 2s3d            & ^3D_3     & 2s2p      & ^3P_2     \\
11 & 49.217 & 1.40(8)  & 49.190 & 1.27(6)  & 1.44(14) & 49.22 & 118& {\rm Si~XI  } & 49.222 & 1.06 & 94 & 2s3d            & ^1D_2     & 2s2p      & ^1P_1     \\
12 & 49.700 & 0.22(7)  & 49.673 & 0.32(6)  & 0.29(7)  & 49.71 & 43 & {\rm Si~X   } & 49.701 & 0.09 & 1  & 2s2p(^1P)3d     & ^2F_{7/2} & 2s2p^2    & ^2D_{5/2} \\
13 & 50.361 & 0.42(10) & 50.334 & 0.44(8)  & 0.51(8)  & 50.35 & 195& {\rm Si~X   } & 50.333 & 0.20 & 1  & 2s2p(^3P)3d     & ^4D_{7/2} & 2s2p^2    & ^4P_{5/2} \\
14 & 50.525 & 1.61(10) & 50.512 & 1.38(9)  & 1.68(15) & 50.52 & 12 & {\rm Si~X   } & 50.524 & 1.45 & 12 & 2s^23d          & ^2D_{3/2} & 2s^22p    & ^2P_{1/2} \\
15a& 50.692 & 1.43(10) & 50.676 & 1.04(8)  & 1.30(14) & 50.69 & 74 & {\rm Si~X   } & 50.691 & 1.08 & 19 & 2s^23d          & ^2D_{5/2} & 2s^22p    & ^2P_{3/2} \\
15b&        &          &        &          &          &       &    & {\rm Si~X   } & 50.703 & 0.29 & 2  & 2s^23d          & ^2D_{3/2} & 2s^22p    & ^2P_{3/2} \\
16 & {\bf 50.874} & 0.23(9)  & 50.828 & 0.15(8)  &          &       &    & {\rm Si~X   } & 50.824 & 0.20 & 0  & 2s2p(^3P)3d     & ^4F_{3/2} & 2s2p^2    & ^4P_{1/2} \\
17 & 52.307 & 0.87(7)  & 52.304 & 0.88(13) & 0.75(11) & 52.30 & 88 & {\rm Si~XI  } & 52.298 & 0.94 & 85 & 2s3s            & ^1S_0     & 2s2p      & ^1P_1     \\
18 & {\bf 52.453} & 0.29(7)  & 52.473 & 0.40(11) &          & 52.48 & 15 & {\rm Si~X   } & 52.484 & 0.13 & 3  & 2s2p(^3P)3d     & ^2F_{7/2} & 2s2p^2    & ^2D_{5/2} \\
19 & 52.594 & 0.30(7)  &        &          & 0.35(8)  & 52.61 & 40 & {\rm Si~X   } & 52.612 & 0.33 & 3  & 2s2p(^3P)3d     & ^2F_{5/2} & 2s2p^2    & ^2D_{3/2} \\
20a&        &          & 55.078 & 0.67(7)  & 0.68(15) & 55.11 & 17 & {\rm Si~IX  } & 55.094 & 0.15 & 6  & 2s^22p3d        & ^3P_0     & 2s^22p^2  & ^3P_1     \\
20b&        &          &        &          &          &       &    & {\rm Si~IX  } & 55.116 & 0.34 & 14 & 2s^22p3d        & ^3P_1     & 2s^22p^2  & ^3P_1     \\
21a&        &          & 55.246 & 1.19(8)  & 0.88(25) & 55.23 & <10& {\rm Si~IX  } & 55.234 & 0.31 & 13 & 2s^22p3d        & ^3P_1     & 2s^22p^2  & ^3P_2     \\
21b&        &          &        &          &          & 55.28 & 28 & {\rm Si~IX  } & 55.272 & 0.51 & 21 & 2s^22p3d        & ^3P_2     & 2s^22p^2  & ^3P_2     \\
22a&        &          & 55.347 & 2.26(9)  & 2.14(27) & 55.37 & 22 & {\rm Si~IX  } & 55.356 & 0.97 & 39 & 2s^22p3d        & ^3D_2     & 2s^22p^2  & ^3P_1     \\
22b&        &          &        &          &          &       &    & {\rm Si~IX  } & 55.383 & 0.20 & 8  & 2s^22p3d        & ^3D_1     & 2s^22p^2  & ^3P_1     \\
22c&        &          &        &          &          & 55.41 & 43 & {\rm Si~IX  } & 55.401 & 1.06 & 43 & 2s^22p3d        & ^3D_3     & 2s^22p^2  & ^3P_2     \\
23 &        &          & 56.017 & 0.56(20) & 0.19(11) & 56.04 & 20 & {\rm Si~IX  } & 56.027 & 0.07 & 3  & 2s2p^2(^2S)3p   & ^3P_0     & 2s^22p^2  & ^1D_2     \\
24 &        &          & 57.196 & 0.25(7)  &          & 57.20 & 25 & {\rm Si~X   } & 57.208 & 0.29 & 6  & 2s2p(^3P)3s     & ^2P_{3/2} & 2s2p^2    & ^2D_{5/2} \\
25 &        &          & 57.309 & 0.35(8)  &          & 57.35 & 17 & {\rm Si~X   } & 57.366 & 0.62 & 4  & 2s2p(^3P)3s     & ^2P_{3/2} & 2s2p^2    & ^2D_{3/2} \\
26a& 61.012 & 1.66(14) &        &          & 1.41(25) & 61.03 & 17 & {\rm Si~VIII} & 60.989 & 0.58 & 8  & 2s^22p^2(^3P)3d & ^4P_{1/2} & 2s^22p^3  & ^4S_{3/2} \\
26b&        &          &        &          &          &       &    & {\rm Si~VIII} & 61.022 & 1.14 & 17 & 2s^22p^2(^3P)3d & ^4P_{3/2} & 2s^22p^3  & ^4S_{3/2} \\
27 & 61.090 & 1.53(14) &        &          & 1.38(24) & 61.09 & 24 & {\rm Si~VIII} & 61.032 & 1.65 & 24 & 2s^22p^2(^3P)3d & ^4P_{5/2} & 2s^22p^3  & ^4S_{3/2} \\
28 & 61.611 & 0.79(13) &        &          & 0.52(17) & 61.61 & 18 & {\rm Si~IX  } & 61.600 & 0.07 & 3  & 2s^22p3s        & ^3P_1     & 2s^22p^2  & ^3P_0     \\
29 & 61.847 & 0.85(13) &        &          & 0.67(11) & 61.85 & 21 & {\rm Si~IX  } & 61.844 & 0.08 & 3  & 2s^22p3s        & ^3P_1     & 2s^22p^2  & ^3P_1     \\
30 & 61.937 & 0.89(13) & 61.971 & 1.11(14) & 0.55(17) & 61.92 & 32 & {\rm Si~VIII} & 61.792 & 0.18 & 3  & 2s^22p^2(^1D)3d & ^2D_{5/2} & 2s^22p^3  & ^2D_{3/2} \\
31 & 63.715 & 0.91(15) & 63.777 & 0.57(11) & 0.58(11) & 63.72 & 352& {\rm Si~VIII} & 63.716 & 0.17 & 2  & 2s^22p^2(^3P)3d & ^2F_{7/2} & 2s^22p^3  & ^2D_{5/2} \\
32 &        &          & {\bf 64.668} & 0.27(7)  &          &       &    & {\rm Si~IX  } & 64.815 & 0.19 & 8  & 2s2p^2(^4P)3s   & ^3P_2     & 2s2p^3    & ^3D_3     \\
33 &        &          & {\bf 64.767} & 0.27(7)  &          &       &    & {\rm Si~IX  } & 64.964 & 0.40 & 16 & 2s2p^2(^4P)3s   & ^3P_1     & 2s2p^3    & ^3D_2     \\
34a& 67.152 & 0.73(16) & 67.143 & 0.56(25) & 0.48(11) & 67.15 & 17 & {\rm Si~IX  } & 67.066 & 0.04 & 2  & 2s2p^2(^4P)3s   & ^3P_2     & 2s2p^3    & ^3P_1     \\
34b&        &          &        &          &          &       &    & {\rm Si~IX  } & 67.071 & 0.11 & 4  & 2s2p^2(^4P)3s   & ^3P_2     & 2s2p^3    & ^3P_2     \\
35a& 67.259 & 0.83(16) & 67.287 & 0.58(14) & 0.87(18) & 67.24 & 42 & {\rm Si~IX  } & 67.222 & 0.07 & 3  & 2s2p^2(^2D)3s   & ^3D_1     & 2s2p^3    & ^3P_1     \\
35b&        &          &        &          &          &       &    & {\rm Si~IX  } & 67.224 & 0.10 & 4  & 2s2p^2(^2D)3s   & ^3D_1     & 2s2p^3    & ^3P_0     \\
35c&        &          &        &          &          &       &    & {\rm Si~IX  } & 67.227 & 0.10 & 4  & 2s2p^2(^2D)3s   & ^3D_1     & 2s2p^3    & ^3P_2     \\
36a& 67.417 & 0.72(15) & 67.369 & 0.39(12) & 0.68(14) & 67.38 & 41 & {\rm Si~VIII} & 67.478 & 0.20 & 3  & 2s2p^3(^5S)3d   & ^4D_{1/2} & 2s2p^4    & ^4P_{1/2} \\
36b&        &          &        &          &          &       &    & {\rm Ne~VIII} & 67.382 &      &    & 4p              & ^2P_{3/2} & 2s        & ^2S_{1/2} \\
37 & 68.132 & 0.41(7)  &        &          &          &       &    & {\rm Si~VII } & 68.148 & 0.38 & 39 & 2s^22p^3(^2P)3d & ^3D_3     & 2s^22p^4  & ^3P_2     \\
38a& 69.641 & 2.19(11) & 69.694 & 2.31(14) & 2.03(21) & 69.65 & 256& {\rm Si~VIII} & 69.632 & 0.70 & 10 & 2s^22p^2(^3P)3s & ^4P_{5/2} & 2s^22p^3  & ^4S_{3/2} \\
38b&        &          &        &          &          &       &    & {\rm Si~VII } & 69.664 & 0.46 & 47 & 2s^22p^3(^2D)3d & ^3P_2     & 2s^22p^4  & ^3P_2     \\
39 & 69.797 & 1.22(11) & 69.864 & 1.23(12) & 1.05(14) & 69.84 & 14 & {\rm Si~VIII} & 69.790 & 0.38 & 6  & 2s^22p^2(^3P)3s & ^4P_{3/2} & 2s^22p^3  & ^4S_{3/2} \\
40a& {\bf 69.909} & 0.67(9)  & 69.993 & 0.73(12) &          & 69.89 & 14 & {\rm Si~IX  } & 69.896 & 0.46 & 19 & 2s^22p3p        & ^3D_3     & 2s2p^3    & ^3D_2     \\
40b&        &          &        &          &          &       &    & {\rm Si~VIII} & 69.905 & 0.20 & 3  & 2s^22p^2(^3P)3s & ^4P_{1/2} & 2s^22p^3  & ^4S_{3/2} \\
41 & 70.050 & 0.80(9)  & 70.102 & 0.48(12) & 0.70(11) & 70.05 & 62 & {\rm Si~VII } & 70.027 & 0.61 & 62 & 2s^22p^3(^2D)3d & ^3D_3     & 2s^22p^4  & ^3P_2     \\
\hline
         \end{array}
      \]
      \end{table*}
\setcounter{table}{1}
\begin{table*}
\centering
    \caption[I]{\it ---Continued}
    \vspace{-0.2cm}
      \[\hspace{-0.7cm}
      \begin{array}[h]{lccccccclcccllll} \hline\hline
         & & & & & & & & & \multicolumn{3}{c}{\rm This~work} & \multicolumn{4}{c}{\rm Transitions} \\
{\rm ID} & {\rm \lambda_{-}~(\AA)} & {\rm Flux_{-}} & {\rm
\lambda_{+}~(\AA)} & {\rm Flux_{+}} & {\rm RMA02} & {\rm
\lambda_{\sun}} &  F_{\rm \sun} & {\rm Ions} & {\rm \lambda~(\AA)}
& F_{\tiny 3CIE} & F^p_{\tiny \sun}& \multicolumn{2}{c}{\rm
Upper~level} & \multicolumn{2}{c}{\rm Lower~level} \\ \hline
42 & 72.443 & 0.36(15) & 72.469 & 0.17(17) &          & 72.41 & 19 & {\rm Si~VIII} & 72.421 & 0.15 & 2  & 2s^22p^2(^3P)3s & ^2P_{1/2} & 2s^22p^3  & ^2D_{3/2} \\
43 & {\bf 73.189} & 0.40(12) & 73.124 & 0.50(22) &          &       &    & {\rm Si~VII } & 73.123 & 0.64 & 65 & 2s^22p^3(^4S)3d & ^3D_3     & 2s^22p^4  & ^3P_2     \\
44 & {\bf 76.010} & 0.90(18) & 76.038 & 0.56(18) & 0.77(13) & 76.03 & 55 & {\rm Si~VIII} & 75.988 & 0.71 & 10 & 2s2p^3(^5S)3s   & ^4S_{3/2} & 2s2p^4    & ^4P_{5/2} \\
45 & 76.157 & 0.25(15) & 76.186 & 0.71(18) &          & 76.16 & 21 & {\rm Si~VIII} & 76.196 & 0.46 & 7  & 2s2p^3(^5S)3s   & ^4S_{3/2} & 2s2p^4    & ^4P_{3/2} \\
\hline
         \end{array}
      \]
      \end{table*}

\begin{figure*}
\centering
\includegraphics[angle=0,width=14cm,height=10cm]{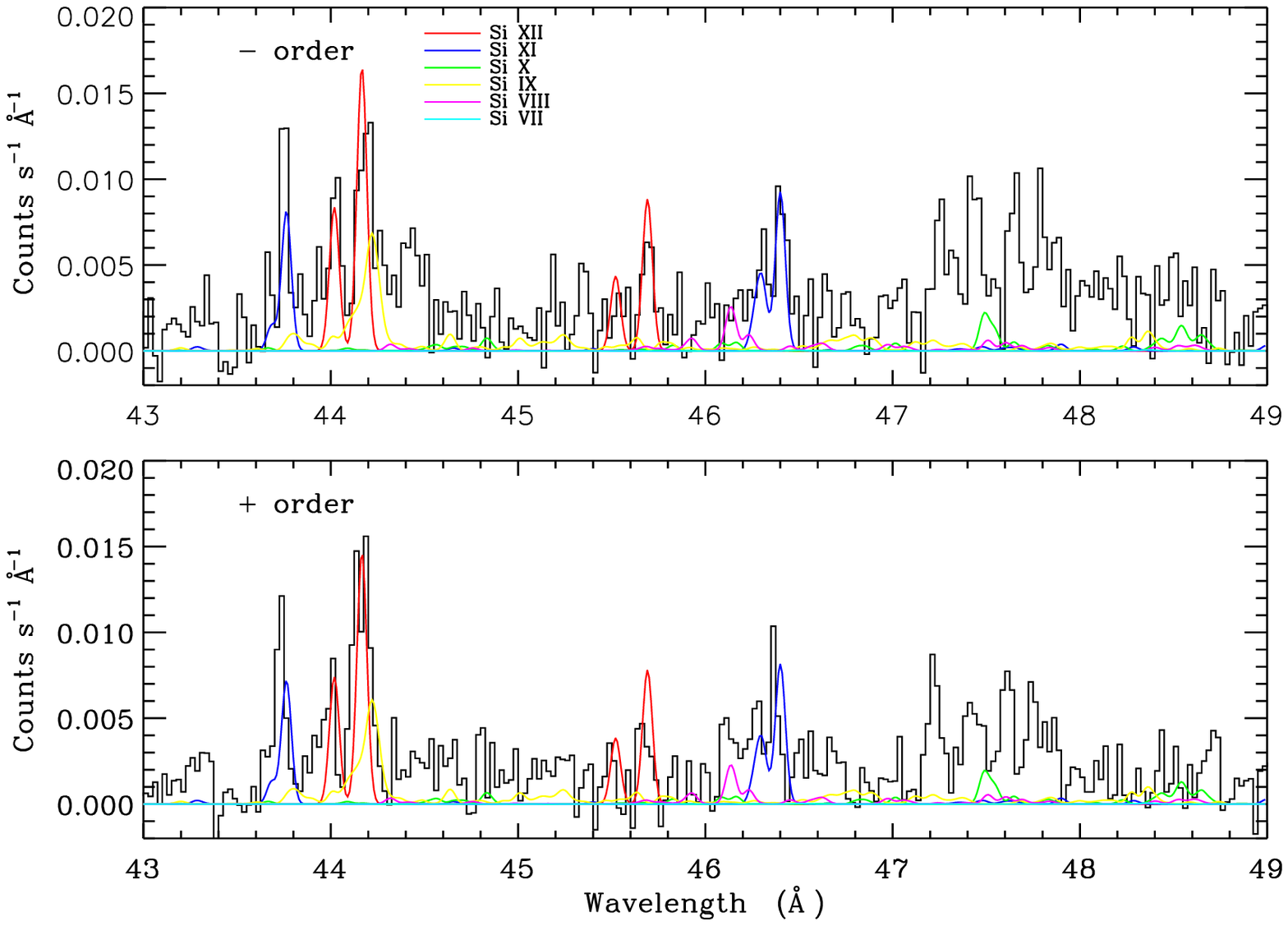}
\includegraphics[angle=0,width=14cm,height=10cm]{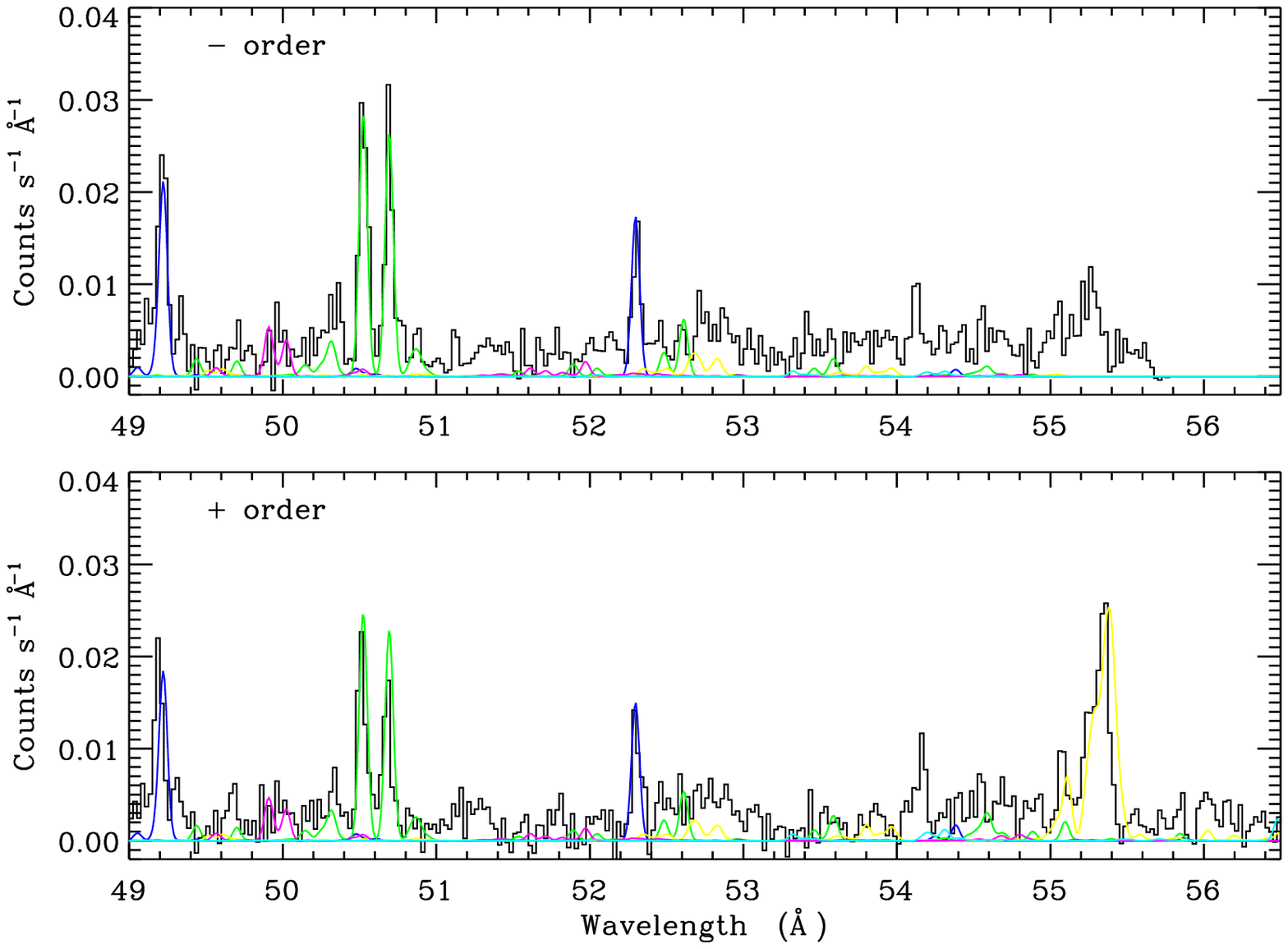}
\caption[short title]{The co-added spectrum (with background
extracted) of different observations (with Obs\_IDs 63, 1461 and
1224) in wavelength range of 43--56~\AA\,, is illustrated by
histogram curves. The positive ({\it bottom}) and negative ({\it
top}) panels are shown separately. The theoretical spectra of
Si~VII---Si~XII are also overlapped with different colors. For
clarity, the spectra in range of 43--88~\AA\, is divided into five
sections as shown in Figs.~2--3. (online color)}
\end{figure*}

{\bf Si~XII}~\hspace{0.5cm} Three emission lines at 44.029, 44.183
and 45.694~\AA\, were identified in work of Raassen et
al.~\shortcite{RMA02}. These lines are reproduced again in present
prediction, and their intensities are slightly higher than the
observations, yet are still within 1$\sigma$ statistical error.
The scaled theoretical line intensities are also in agreement with
the solar observation as shown in Table~2. Furthermore, we notice
an emission at 45.550~\AA\,, its intensity agrees with the Procyon
observation in the both diffractions, as shown in Fig.~1. So we
assign the emission to the $3s~^2S_{1/2}$--$2p~^2P_{1/2}$
transition with wavelength of 45.521~\AA\,. For the stellar
spectra, this assignment is the first time to our best knowledge.
This assignment is confirmed in solar observation with higher
resolution (0.02~\AA\,).

{\bf Si~XI}~\hspace{0.5cm} The present prediction satisfactorily
reproduces the emission lines at 43.753, 46.301, 46.407, 49.217
and 52.307~\AA\, as illustrated in Table 2 and Fig.~1. Around the
peak at 43.753~\AA\,, a weak line with wavelength of 43.663~\AA\,
is predicted, which can explain the left-side wing of the emission
at 43.753~\AA\,. In solar flare observation, this line is clearly
resolved with a wavelength of 43.65~\AA\,. Yet it still has not
been identified so far. For the emission line at 52.307~\AA\,, the
calculation agrees with the observed flux in both diffractions, as
shown in Table 2 and Fig.~1. However, Acton et
al.~\shortcite{ABB85} pointed that there is a contamination from
Al~XI line at 52.244~\AA\,. Present prediction indicates that the
contribution from Al~XI line can be negligible for Procyon and
solar.

The line ratios among these emission lines show powerful
diagnostic potentials for the electron density and temperature as
revealed in work of Liang \& Zhao~\shortcite{LZ06}. For example,
the intensity of 43.753~\AA\, line is sensitive to the electron
temperature, and increases relative to the intensity of the
52.307~\AA\, line with increasing the electron temperature [see
Fig.~2 in Ref. Liang \& Zhao~\shortcite{LZ06}]. The larger
discrepancy ($\sim$50\%) of this line between the prediction and
solar value reveals the higher electron temperature in solar
flare.

{\bf Si~X}~\hspace{0.5cm} The two strong emission lines at 50.525
and 50.692~\AA\, show an excellent $n_{\rm e}-$diagnostic
potential for inactive cool stars as reported in our previous
work~\cite{LZS06}. They are satisfactorily reproduced at the
electron density of 3.0$\times10^8$~cm$^{-3}$ and the EM
distribution determined by Raassen et al.~\shortcite{RMA02}. For
the emission at 50.361~\AA\,, Raassen et al.~\shortcite{RMA02}
assigned it to Si~X according to the line-list of Kelly
database~\cite{Kel87}\footnote{http://cfa-www.harvard.edu/amp/ampdata/kelly/kelly.html},
whereas the database is based upon the high-density plasma such as
laser plasma. Fortunately, present work also predicts the emission
with wavelength of 50.333~\AA\,, but it contributes about
$\sim$48\% in both diffractions. This means that there is unknown
contamination from other lines. In solar flare observation, Acton
et al.~\shortcite{ABB85} assign the observed emission at
50.35~\AA\, to Fe~XVI line with wavelength of 50.350~\AA\,. So the
left contribution at 50.361~\AA\, maybe from the Fe~XVI in Procyon
coronae. However, In solar flare, the prediction reveals that the
contribution from Si~X at 50.35~\AA\, of spectrum can be
negligible. For peaks at 50.874 and 52.594~\AA\,, the predicted
line fluxes from the $2s2p(^3P)3d~^4F_{3/2}$--$2s2p^2~^4P_{1/2}$
(50.824~\AA\,) and $2s2p(^3P)3d~^2F_{5/2}$--$2s2p^2~^2D_{3/2}$
(52.612~\AA\,) transitions are comparable with the observed values
within 1$\sigma$ error, which indicates the contamination from
Ni~XVIII as revealed by Raassen et al.~\shortcite{RMA02} is
negligible. But, in solar flare, the contribution from Ni~XVIII is
dominant, because Si~X contributes only $\sim$8\%. The present
calculation predicts another emission lines at 52.484~\AA\,.
Moreover it intensity is 33\% and 43\% when compared to the
observed flux in negative and positive diffractions, respectively.
In solar flare, the contribution from Si~X is less than 7\% at
52.48~\AA\, that is the emission line is still an unknown line. At
the emission of 49.700~\AA\,, the contribution from Si~X
(49.701~\AA\,) is predicted to be $\sim$36\% and 2\%, in Procyon
and Sun, respectively.

In wavelength range of 57.0---57.5~\AA\,, two emission lines
($\lambda=$57.196 and 57.309~\AA\,) are detected in the positive
spectrum, whereas they haven't been identified so far. In negative
order spectrum, there are no counts between 55.7--58.5~\AA\,,
because of the chip gap. Present work predicts line flux at
wavelengths of 57.208~\AA\, being consistent with the observed
values at 57.196, while the flux at 57.366 being higher than
observed value at 57.309~\AA\, by 75\%. So we tentatively assign
the two emission lines to the
$2s2p(^3P)3s~^2P_{3/2}$~--~$2s2p^2~^2D_{5/2}$ and
$2s2p(^3P)3s~^2P_{1/2}$~--~$2s2p^2~^2D_{3/2}$ transitions of Si~X.
In solar flare, 24\% contribution of Si~X is predicted for the two
emissions.
\begin{figure*}
\centering
\includegraphics[angle=0,width=14cm,height=10cm]{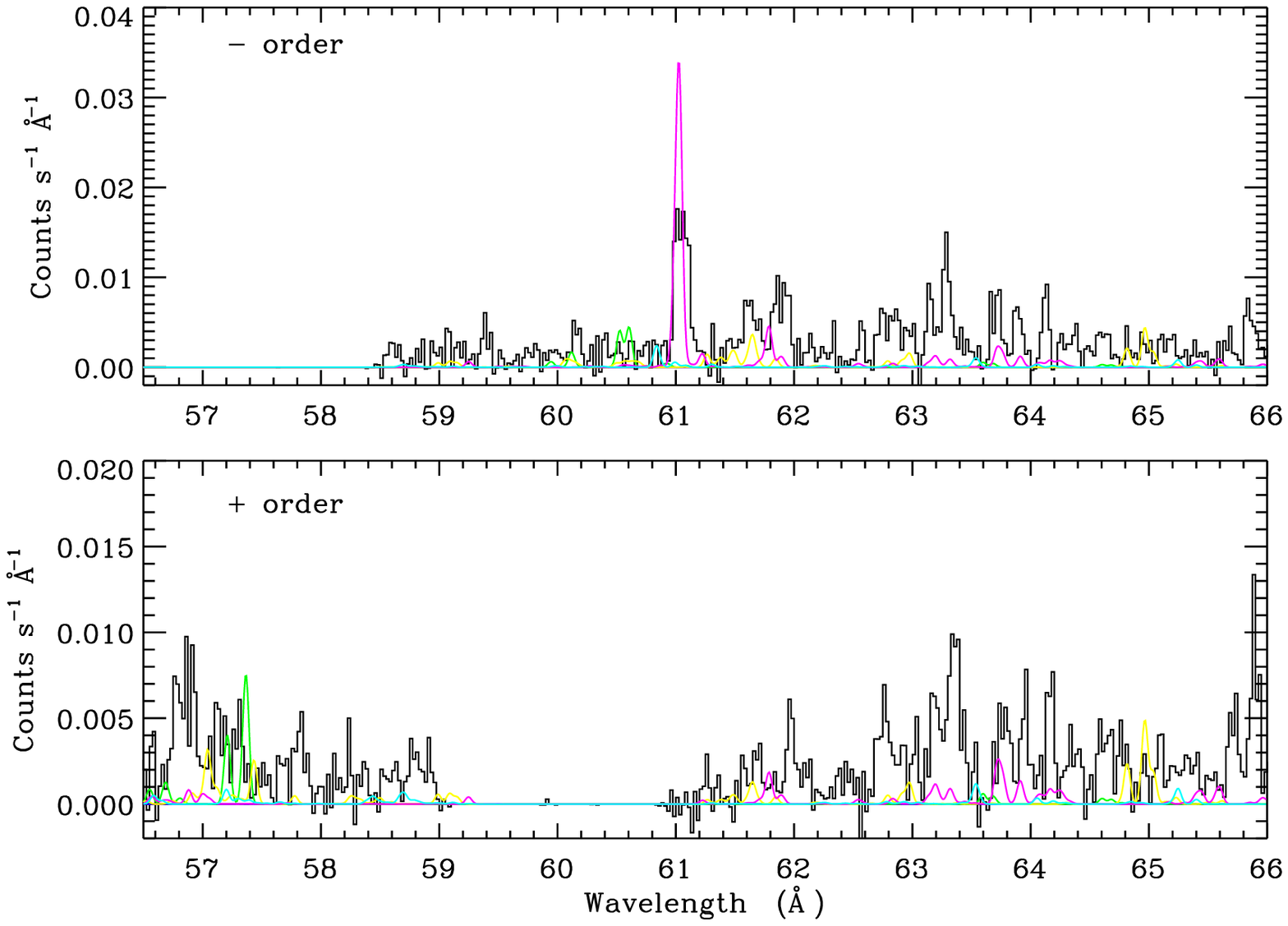}
\includegraphics[angle=0,width=14cm,height=10cm]{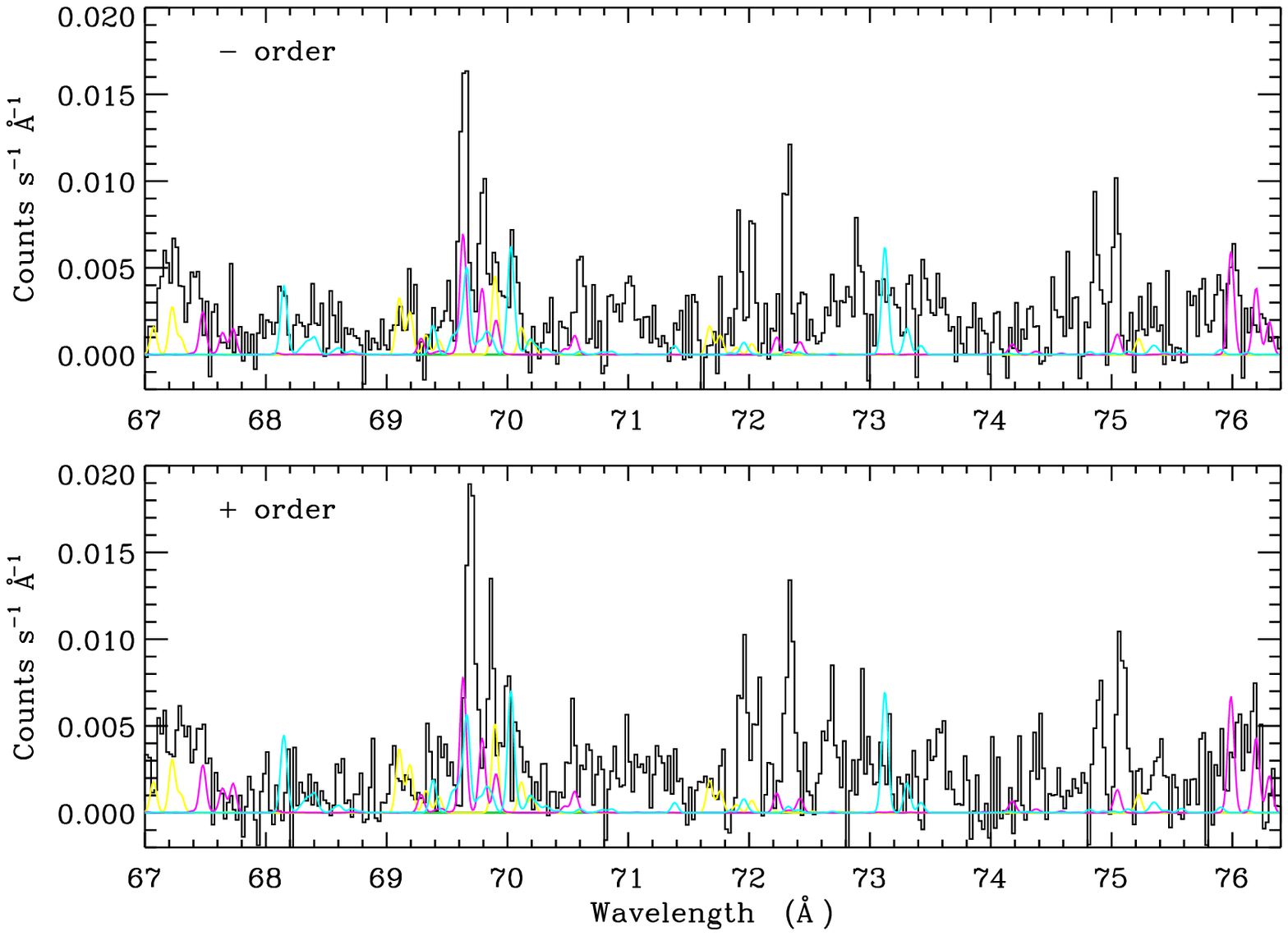}
\caption[short title]{The co-added spectrum with background
extracted in wavelength range of 56.5--76.4~\AA\,. The description
can refers to the caption of Fig.~1. (online color)}
\end{figure*}

{\bf Si~IX}~\hspace{0.5cm} Present calculation satisfactorily
reproduces the line emissions at 55.246 and 55.347~\AA\,. The line
intensity ratio between them is sensitive to the electron density
($n_{\rm e}$), whereas insensitive to the electron temperature
($T_{\rm e}$) with variation less than 3\% over log$T_{\rm
e}$~(K)=5.9--6.3 [see Fig.~3 in Ref. Liang \&
Zhao~\shortcite{LZ07}]. The contribution of Si~IX to the observed
flux at 55.078~\AA\, occupies a considerable part. When 14\%
contribution from Si~X is considered, the contamination from Mg~IX
pointed out by Raassen et al.~\shortcite{RMA02} can be negligible.
At the peak at 56.017~\AA\,, we found the contribution from Si~IX
is very small. The major component maybe from S~IX (56.081~\AA\,)
and Ni~XIII (56.000~\AA\,) as pointed out by Raassen et
al.~\shortcite{RMA02}. In solar flare, the contribution from Si~IX
(56.027~\AA\,) is similarly very small ($\sim$15\%).  At the
position about 44.218~\AA\,, a large difference is found between
the both diffractions, which is due to the low statistical
performance in the interested wavelength region. The contribution
from Si~IX is estimated to be $\sim$100\% and 50\% in the negative
and positive diffractions, respectively. Additionally,
contamination from second-order diffraction of O~VII line
(2$\times$22.0975~\AA\,) may be the origin of the difference, as
pointed out by Acton et al.~\shortcite{ABB85} in the observation
of solar flare.

\begin{figure*}
\centering\vspace{-0.6cm}
\includegraphics[angle=0,width=14cm,height=10cm]{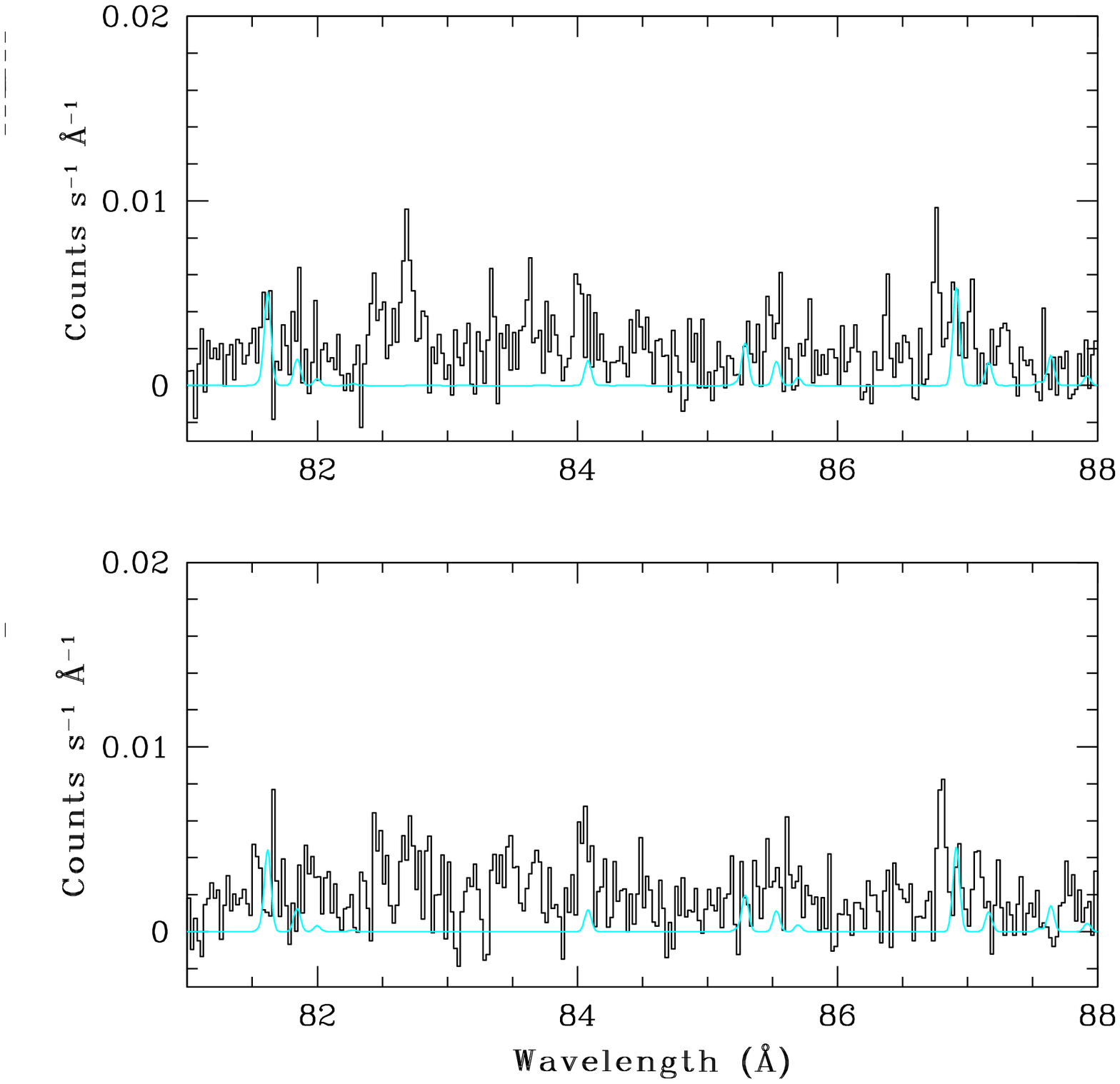}
\caption[short title]{The co-added spectrum with background
extracted in wavelength range of 81--88~\AA\,. The description can
refers to the caption of Fig.~1. (online color)}
\end{figure*}
The two $3s$--$2p$ transition lines at 61.611 and 61.847~\AA\, are
underestimated greatly in Procyon and solar flare. The large
discrepancies compared with the $3d$--$2p$ lines at 55.359~\AA\,,
maybe from other unknown contamination. Indirect processes for
populations on $3s$ and $3d$ levels and opacity effects also maybe
the reasons of the deviations as in the debated case of
Fe~XVII~\cite[ect.]{NSA03,BBG04}. The theoretical calculation also
predicts an considerable line fluxes at 64.815 and 64.964~\AA\,
being from $2s2p^2(^4P)3s~^3P_2$--$2s2p^3~^3D_3$ and
$2s2p^2(^2D)3s~^3P_1$--$2s2p^3~^3D_2$ transitions. But, no
emission lines are detected at the two positions. We notice that
the predicted fluxes agree with the observed values at measured
wavelength of 64.668 and 64.767~\AA\,. So we tentatively assign
the two emission lines from $3s$--$2p$ transitions of Si~IX by
consideration of the 1\% uncertainty of level energies for double
excited levels.

Though the lines at 67.152 and 67.259~\AA\, were assigned to be
Mg~IX (67.132~\AA\,) and Ne~VIII (67.350~\AA\,), respectively, by
Raassen et al.~\shortcite{RMA02}, we also notice the contamination
($\sim$20--40\%) from Si~IX in Procyon and solar flare. The
contribution of Si~IX (69.896~\AA\,) is up to $\sim$70\% at the
peak of 69.909~\AA\, as shown in Fig.~2. The left contribution is
from Si~VIII (69.905~\AA\,). In solar flare, the contribution from
Si~VIII is predicted to be less than 20\%.

{\bf Si~VIII}~\hspace{0.5cm} The line ratio between $3d$--$2p$
(61.022~\AA\,) and $3s$--$2p$ (69.646~\AA\,) lines is predicted
higher than observation again as revealed in Si~IX~\cite{LZ07} and
Fe~XVII~\cite[and references therein]{DB02}. The predicted line
fluxes at 61.012 and 61.090~\AA\, are slightly overestimated, but
agree with observations within 1$\sigma$ error for Procyon and
solar flare observations. This reveals that the contamination from
Mg~IX (61.088~\AA\,) can be negligible. At peaks of 69.641 and
69.797~\AA\,, the contribution from $3s$--$2p$ transition of
Si~VIII is about 32\%. Table~2 indicates that the line intensity
of Si~VII line (69.664~\AA\,) is $\sim$65\%  relative to the
intensity of Si~VIII line (69.632~\AA\,). The total contribution
from Si~VIII (69.632~\AA\,) and Si~VII (69.664~\AA\,) are
predicted to be $\sim$50\% and $\sim$25\% for Procyon and solar
flare, that is there is an unknown contamination besides the minor
contribution of Fe~XIV as reported by Acton et
al.~\shortcite{ABB85}.

We also notice that the prediction at 75.988 and 76.196~\AA\,
agree well with the observed fluxes at 76.010 and 76.157~\AA\,.
When a deviation of 1\% in the double excited energy levels is
considered, we tentatively assign the two emissions to Si~VIII
lines with wavelengths of 75.988 and 76.196~\AA\,. In solar flare,
the contribution from Si~VIII is estimated to be $\sim$20--40\%.
Acton et al.~\shortcite{ABB85} assign the two emission lines to
Fe~XIV (76.023~\AA\,) and Fe~XIII (76.152~\AA\,), respectively.
Moreover, 10~photons$\cdot$cm$^{-2}$s$^{-1}\cdot$acrsec$^{-1}$ at
68.85~\AA\, is detected in solar flare, which was assigned to the
$2p^23s~^4P_{3/2}$--$2p^3~^4S_{3/2}$ transition of Si~VIII with
wavelength of 68.853~\AA\, by Acton et al.~\shortcite{ABB85}. Yet,
no dominant flux of Si~VIII is predicted at this wavelength
region. The lower contributions from Si~VIII for solar flare, is
due to the solar flare is hotter than Procyon coronae.

{\bf Si~VII}~\hspace{0.5cm} As illustrated in Fig.~3 and Table 2,
the line flux of Si~VII (70.027~\AA\,) occupies about nearly 80\%,
that is the contribution from Fe~XII (70.010~\AA\,) and Fe~XV
(70.054~\AA\,) is very small. Moreover, in the positive spectrum,
the observed wavelengths around this region systematically longer
than calculations and the observed values in the negative spectrum
by $\sim$0.06~\AA\,. In solar flare, the contribution from Fe~XV
become strong as the assignment by Acton et al.~\shortcite{ABB85}.
At the peak around 68.132~\AA\, of the negative spectrum of
Procyon, the predicted intensity agrees well with observed value.
So we tentatively assign the emission line to
$2s^22p^3(^2P)3d~^3D_3$--$2s^22p^4~^3P_2$ transition
(68.148~\AA\,) of Si~VII. This assignment is the first
identification for stellar coronal spectra.

At peaks of 73.189, 81.61 and 86.876~\AA\,, the calculation
satisfactorily reproduces the Procyon observations. However, the
former two lines haven't been identified so far, and the last
emission line is assigned to Mg~VIII by according to the
line-lists of Kelly database. In the low-density astrophysical
plasma, the contribution from Mg~VIII can be negligible. At the
peak of 81.61~\AA\, in the positive spectrum, the observed
wavelength is shorter than that in negative spectrum and the
calculation by $\sim$0.06~\AA\,. This maybe due to the wavelength
calibration in the positive diffraction.

\section{Conclusion}
In summary, available observation data (Obs\_IDs of 63, 1461 and
1224) for cool star--Procyon are co-added and analyzed. By
adopting our published atomic data and some unpublished data, line
emissivities of Si~VI--Si~XII ions are calculated at an electron
density of 3.0$\times10^8$~cm$^{-3}$ (typical value for cool
stars) and temperatures over 0.1--5.0~MK. In this work, energy
levels of 878, 312, 560, 320, 350 and 40 have been included for
Si~VII--Si~XII, respectively. Moreover, decay rates of transitions
(including E1, M1, E2 and M2) up to tens of thousand are included.
Some electron impact excitation data are replaced by available
$R-$matrix data to take resonant effects into account as far as
possible.

Based upon the EM derived by Raassen et al.~\shortcite{RMA02} and
the calculated emissivities, we estimated the theoretical line
fluxes of highly charged Si~VII--Si~XII ions. By detailed
comparison between the observations and predictions, several
emissions lines [with bold font in second or forth column of Table
2] are identified firstly to our best knowledge. The
identification is assessed by comparing with solar spectrum with
higher resolution (0.02~\AA\,). The prediction indicates the large
discrepancies between the $3s$--$2p$ line (69.641~\AA\,) {\it
versus} $3d$--$2p$ line (61.012~\AA\,) again for lower charge
stage Si~VIII. This reveals that more accurate calculation of
cross section in electron-ion interactions is very necessary. For
emission lines at 52.594~\AA\, (Si~X), 55.078~\AA\, (Si~IX) and
70.027 (Si~VII), this work indicates that the contamination from
Ni~XVIII (52.615~\AA\,), Mg~IX (55.060~\AA\,) and Fe~XII (70.010
and 70.054~\AA\, of Fe~XV), respectively, can be negligible for
Procyon. However, in solar flare, the contaminations from Mg, Fe
and Ni ions become dominant. This is due to the solar flare is
hotter than the Procyon coronae. For the peak around 69.641~\AA\,,
Si~VII line has the comparable contribution with the Si~VIII lines
identified by Raassen et al.~\shortcite{RMA02}. However, their
total contribution is less than 25\% in solar flare. The
discrepancies and the identification strongly suggest the
benchmark from laboratory measurements at low-density plasmas.

\section*{Acknowledgments}
This work was supported by the National Natural Science Foundation
of China under grant Nos. 10603007 and 10521001, as well as
National Basic Research Program of China (973 Program) under grant
No. 2007CB815103.

\bsp \label{lastpage}
\end{document}